\documentclass[12pt]{article}

\usepackage{amssymb}
\usepackage{amsmath}
\usepackage{amscd}
\usepackage{latexsym}
\usepackage{graphicx}

\usepackage{caption}

\usepackage{cite}

\newcommand{\be}{\begin{equation}}
\newcommand{\ee}{\end{equation}}

\newcommand{\dlt}{\delta}
\newcommand{\prt}{\partial}
\newcommand{\br}{{\bf r}}
\newcommand{\ba}{{\bf a}}

\newcommand{\bt}{\beta}
\newcommand{\vp}{\varphi}

\newcommand{\al}{\alpha}
\newcommand{\ra}{\rightarrow}
\newcommand{\sgm}{\sigma}

\newcommand{\gm}{\gamma}

\newcommand{\Om}{\Omega}

\newcommand{\dgr}{\dagger}
\newcommand{\lbd}{\lambda}
\newcommand{\Lbd}{\Lambda}
\newcommand{\rgl}{\rangle}
\newcommand{\lgl}{\langle}
\newcommand{\cH}{{\cal H}}

\newcommand{\cD}{{\cal D}}

\begin{document}

\begin{center}

{\Large{\bf Statistical model of a superfluid solid} \\ [5mm]

V.I. Yukalov$^{1,2}$ and E.P. Yukalova$^{3}$ }  \\ [3mm]

{\it
$^1$Bogolubov Laboratory of Theoretical Physics, \\
Joint Institute for Nuclear Research, Dubna 141980, Russia \\ [2mm]

$^2$Instituto de Fisica de S\~ao Carlos, Universidade de S\~ao Paulo, \\
CP 369, S\~ao Carlos 13560-970, S\~ao Paulo, Brazil \\ [2mm]

$^3$Laboratory of Information Technologies, \\
Joint Institute for Nuclear Research, Dubna 141980, Russia } \\ [3mm]

{\bf E-mails}: {\it yukalov@theor.jinr.ru}, ~~ {\it yukalova@theor.jinr.ru}

\end{center}

\vskip 1cm

\begin{abstract}

A microscopic statistical model of a quantum solid is developed, where inside a 
crystalline lattice there can exist regions of disorder, such as dislocation
networks or grain boundaries. The cores of these regions of disorder are allowed 
for exhibiting fluid-like properties. If the solid is composed of Bose atoms, then 
the fluid-like aggregations inside the regions of disorder can exhibit Bose-Einstein 
condensation and hence superfluidity. The regions of disorder are randomly distributed
throughout the sample, so that for describing the overall properties of the solid 
requires to accomplish averaging over the disordered aggregation configurations. 
The averaging procedure results in a renormalized Hamiltonian of a solid that can 
combine the properties of a crystal and superfluidity. The possibility of such a 
combination depends on the system parameters. In general, there exists a range of the 
model parameters allowing for the occurrence of superfluidity inside the disordered 
aggregations. This microscopic statistical model gives the opportunity to answer which 
real quantum crystals can exhibit the property of superfluidity and which cannot.       

\end{abstract}

\vskip 2mm

{\it Keywords}: Statistical model, Quantum crystals, Regions of disorder, 
Dislocation networks, Superfluidity 

\newpage

\section{Introduction}

The possibility for the existence of solids that could exhibit the effect of 
superfluidity has been a rather hot topic in recent years. Often, one defines
a superfluid solid as a statistical system where simultaneously there occurs
translational as well as gauge symmetry breaking. There are many experimental
\cite{Botcher_61,Chomaz_62,Tanzi_63,Guo_64,Bottcher_45} and theoretical 
\cite{Lu_65,Young_66} works confirming the existence of this double symmetry 
breaking in trapped gases of atoms with dipolar interactions. Similar structures 
are predicted for spin-orbit coupled condensed gases \cite{Adhikari_67,Kaur_68}. 
These, however, are anyway gases but not solids. In addition, the dipolar periodic 
structures are not absolutely stable, since their lifetime is limited by fast 
inelastic losses caused by three-body collisions. Thus, the droplet structure 
of $^{164}$Dy survives for $0.1$ s and of $^{166}$Er, for only $0.01$ s. Here 
we aim at discussing real solid statistical systems, but not spatially modulated 
metastable gases.  

There is a number of publications on the problem of the possible appearance of 
superfluidity in real quantum crystals, such as solid $^4$He, as can be inferred 
from the review articles 
\cite{Prokofiev_1,Boninsegni_3,Chan_4,Hallock_5,Kuklov_6,Yukalov_7,Fil_8}. 
Ideal crystals cannot possess the property of superfluidity \cite{Penrose_9}, but 
some kind of local disorder, such as dislocations or grain boundaries, is compulsory 
\cite{Kuklov_6,Yukalov_7,Fil_8}. Although dislocations and grain boundaries look as 
one-dimensional and two-dimensional objects, respectively, they, strictly speaking, 
are rather quasi-one-dimensional and quasi-two-dimensional, since their cores are of 
nanosize thickness \cite{Hirth_12,Hull_13,Souris_14}. Dislocations can also form 
dislocation networks \cite{Kuklov_6,Fil_8,Shevchenko_15,Fil_16} that could support 
superfluidity.

To describe microscopically a crystal with regions of disorder is a quite difficult 
problem, since such a description has to deal with a nonuniform and, generally, 
nonequilibrium matter, since, e.g., dislocations can move 
\cite{Hirth_12,Hull_13,Souris_14}. Therefore phenomenological models are used 
\cite{Fil_8,Shevchenko_15,Fil_16}. These models are convenient for characterizing 
a matter with presupposed properties, but the weakness of phenomenological models 
is in their inability to answer whether the system does possess these properties, 
for instance whether superfluidity in a crystal with the regions of disorder can 
really arise. 

In the same vein, modeling a system on the basis of a nonlinear Schr\"{o}dinger 
equation cannot prove the possible existence of a superfluid solid. This is 
because employing this equation presupposes that absolutely all particles forming 
the system are Bose-condensed. The description of the system of Bose particles by 
a nonlinear Schr\"odinger equation is nothing but the coherent approximation that 
is valid only for asymptotically weak interactions and zero temperature 
\cite{Bogolubov_17,Bogolubov_18}. Assuming that all particles are Bose-condensed 
implies that the system is $100 \%$ superfluid. This is why describing the system 
by a periodic solution of the nonlinear Schr\"{o}dinger equation, as has been 
suggested by Gross \cite{Gross_19}, can be applicable for weakly interacting Bose 
gases at zero temperature, but not for solids.   

Trapped Bose gases with dipolar interactions can exhibit spatial periodic 
modulation due to the peculiarity of dipolar forces, containing repulsive as well 
as attractive parts, at the same time remaining Bose-condensed, hence superfluid 
\cite{Griesmaier_69,Baranov_70,Ueda_71,Baranov_72,Stamper_73,Gadway_74,Yukalov_75}.

The idea of constructing a microscopic model of a solid with regions of disorder 
that imitate liquid-like properties was advanced in Refs. 
\cite{Yukalov_20,Yukalov_21,Yukalov_22,Yukalov_23}. The possibility that such 
regions of disorder can house Bose-Einstein condensate, hence superfluidity, was 
mentioned in Ref. \cite{Yukalov_24}. 

In the present paper, we develop the ideas of Refs. 
\cite{Yukalov_20,Yukalov_21,Yukalov_22,Yukalov_23,Yukalov_24} and construct 
a microscopic statistical model of a crystal with regions of disorder, where 
superfluidity could exist, provided such a system remains stable. The occurrence 
of superfluidity, of course, depends on the system parameters. Substituting the 
parameters typical of real quantum crystals makes it straightforward to estimate 
whether this material can support superfluidity or not. 
      
The layout of the paper is as follows. In the next Sec. 2, we give the mathematical 
basis required for describing statistical systems that can house several different 
phases. In the following Sec. 3, we specify the consideration to the case of 
coexisting solid-like and liquid-like phases. The last Sec. 4 concludes. 

Everywhere throughout the paper we use the system of units where the Planck and 
Boltzmann constants are set to one.

\section{Mathematical preliminaries}

Here we briefly delineate the main mathematical steps in deriving the models that 
can describe systems where inside one phase there can appear randomly distributed 
regions of other phases. We give just a concise account of the basic points, 
since the details can be found in the review articles \cite{Yukalov_25,Yukalov_26}. 
On the other side, reminding the principal points of the approach seems to be 
important in order that the reader could be convinced that the derivation of the 
effective renormalized Hamiltonian of the model is based on reliable mathematical 
foundations.    

Let us consider a system of $N$ particles (atoms or molecules) in a volume $V$. 
The system can include several thermodynamic phases enumerated by the index 
$f=1,2,\ldots$ so that
\be
\label{1}
 N = \sum_f N_f \; , \qquad  V = \sum_f V_f \; ,
\ee
with $N_f$ and $V_f$ being the number of particles and volume for an $f$-th phase. 
In what follows, for the sake of conciseness, we denote the spatial volume and its 
measure by the same letter. For a while, the nature of the phases is not important.

The regions comprising different phases, as has been explained by Gibbs 
\cite{Gibbs_27}, can be imagined to be separated by an equimolar surface guaranteeing 
the additivity for the number of particles and volume. The spatial location of each 
phase can be characterized \cite{Yukalov_25,Yukalov_26} by the manifold indicator 
functions \cite{Bourbaki_28}
\begin{eqnarray}
\label{2}
\xi_f(\br) = \left\{ \begin{array}{ll}
1 \; , ~ & ~ \br \in V_f \\
0 \; , ~ & ~ \br \not\in V_f
\end{array}  \right.
\end{eqnarray}
satisfying the properties
\be
\label{3}
  \sum_f \xi_f(\br) = 1 \; , \qquad  
\int_V \xi_f(\br) \; d\br = V_f \; .
\ee

In its turn, each volume $V_f$ can be separated into $\nu_f$ smaller cells of 
volumes $V_{fj}$ containing $N_{fj}$ particles, such that
\be
\label{4}
 N_f = \sum_{j=1}^{\nu_f} N_{fj} \; , \qquad  
V_f = \sum_{j=1}^{\nu_f} V_{fj} \; .
\ee
The cell indicator functions (or submanifold indicator functions) are
\begin{eqnarray}
\label{5}
\xi_{fj}(\br) = \left\{ \begin{array}{ll}
1 \; , ~ & ~ \br \in V_{fj} \\
0 \; , ~ & ~ \br \not\in V_{fj} 
\end{array}  \right. 
 \; , 
\end{eqnarray}
with the property
\be
\label{6}
\int_{V_{fj}} \xi_{fj}(\br) \; d\br = V_{fj} \;   .
\ee
The sums of the cell indicator functions compose the manifold indicator functions 
(\ref{2}),
\be
\label{7}
 \xi_f(\br) =  \sum_{j=1}^{\nu_f} \xi_{fj}(\br-\ba_{fj}) \qquad
( \ba_{fj} \in V_{fj} ) \; ,
\ee
where ${\bf a}_{fj}$ is a fixed vector in $V_{fj}$.

With the help of these cells, it is straightforward to characterize the spatial 
regions of any shape. From relations (\ref{4}), it is clear that, when diminishing 
the cell volumes, the cell number increases,
\be
\label{8}
 \nu_f \ra \infty \; , \qquad
V_{fj} \ra 0 \;  .
\ee
The cell indicator functions of different phases are mutually orthogonal. The family 
of the manifold indicator functions (\ref{5}) is an orthogonal set covering $V_f$, 
\be
\label{9}
\xi_f \equiv \{ \xi_{fj}(\br) : ~ \br \in V \; ; ~ j = 1,2,\ldots,\nu_f \} \; .
\ee
The covering set for $V$ is
\be
\label{10}
\xi \equiv \{ \xi_f : ~  f = 1,2,\ldots \} \;   .
\ee
 
The Hilbert space of microscopic states $\cH$ describing a physical system can 
be defined as a closed linear envelope over a basis $\{\vp_n\}$. The average of 
an operator $\hat{A}$ on $\cH$ is
$$
{\rm Tr}_\cH \; \hat\rho \; \hat A = 
\sum_{\vp_n\in\cH} (\vp_n, \; \hat\rho \hat A \vp_n) \;  ,
$$
with $\hat{\rho}$ being the system statistical operator. When a phase transition 
can occur in the system, and some symmetry can be spontaneously broken, the total 
space $\cH$ can be subdivided into subspaces of microscopic states typical 
of particular phases \cite{Ruelle_29,Emch_30,Bratelli_31}. Let $\cH_f\subset\cH$ 
be a space of microscopic states typical of an $f$-th phase. Then the appropriate 
averaging operation, necessary for correctly describing thermodynamic phases, is 
done with the help of the Bogolubov method of quasi-averages 
\cite{Bogolubov_17,Bogolubov_18}, whose essence is equivalent to the Brout method 
of restricted trace \cite{Brout_32}. In these methods, the averaging operation is 
accomplished not over the total space of microstates $\cH$ but over the restricted 
space $\cH_f$, composed of the microstates typical of an $f$-th phase,   
$$
{\rm Tr}_{\cH_f} \; \hat\rho \; \hat A = 
\sum_{\vp_n\in\cH_f} (\vp_n, \; \hat\rho \hat A \vp_n) \;    .
$$
A generalization of the method of restricted trace can be done \cite{Yukalov_25} 
by associating $\cH_f$ with a weighted Hilbert space. This is the standard 
procedure for characterizing pure states of a system, when the whole system is 
in one of the pure phases. 

When the considered system is heterophase, so that some spatial parts of the 
system are in one thermodynamic phase and some are in other phases, the situation 
is quite different. We mean here that the system parts are not macroscopic, but 
rather are mesoscopic, at least in one direction. The term mesoscopic implies that 
the linear size of a phase embryo, at least in one direction, is such that it is 
much larger than the mean interparticle distance but much smaller than the system 
linear size. The phase embryos are assumed to be randomly distributed over the 
system volume. Then the system space of microstates is the tensor product
\be
\label{11}
\widetilde\cH = \bigotimes_f \cH_f \;   .
\ee
The representation of an operator $\hat{A}$ on $\mathcal{H}_f$, associated with 
the region labeled by $\xi_f$, is denoted as $\hat{A}_f(\xi_f)$. The operators of 
observables on space (\ref{11}) have the structure
\be
\label{12}
 \hat A(\xi) = \bigoplus_f \hat A_f(\xi_f) \;  .
\ee
      
The statistical operator $\hat{\rho}(\xi)$ of a heterophase system can be found 
by minimizing an information functional. The statistical operator has to be 
normalized,
\be
\label{13}
{\rm Tr} \int \hat\rho(\xi) \;\cD\xi = 1 \; ,
\ee
where $\cD\xi$ implies a differential measure over randomly distributed phase 
configurations. The trace operation, here and in what follows, if the space is 
not explicitly shown, is over space (\ref{11}). The average energy is given by 
the expression
\be
\label{14}
{\rm Tr} \int \hat\rho(\xi) \hat H(\xi) \; \cD\xi = E \; ,
\ee
where $\hat{H}(\xi)$ is a system energy operator. Also, there may exist other 
constraints that can be written as 
\be
\label{15}
{\rm Tr} \int \hat\rho(\xi) \hat C_i(\xi) \; \cD\xi = C_i 
\qquad ( i = 1, 2, \ldots ) \; .
\ee

The information functional in the Kullback-Leibler form 
\cite{Kullback_33,Kullback_34} writes as
$$
I[\; \hat\rho\; ] = {\rm Tr} \int \hat\rho(\xi) \; 
\ln \; \frac{\hat\rho(\xi)}{\hat\rho_0(\xi)}\; \cD\xi +
\al \left[  {\rm Tr} \int \hat\rho(\xi) \; \cD\xi -1 \right] +
$$
\be
\label{16}
+ 
\bt \left[  {\rm Tr} \int \hat\rho(\xi) \hat H(\xi) \; \cD\xi - E \right]
+  \sum_i \gm_i \left[  {\rm Tr} \int \hat\rho(\xi) \hat C_i(\xi) \; \cD\xi - 
C_i \right] \; ,
\ee
with the Lagrange multipliers $\al$, $\bt$, and $\gm_i$. The multiplier $\bt=1/T$
is the inverse temperature and the multipliers $\gm_i=-\bt\mu_i$ are expressed
through chemical potentials $\mu_i$. The trial statistical operator $\hat\rho_0(\xi)$ 
takes into account additional imposed constraints, if any are known. In the case of 
no imposed trial constraints, the operator $\hat\rho_0(\xi)$ reduces to a constant. 
Then the minimization of the information functional (\ref{16}) yields the statistical 
operator
\be
\label{17}
 \hat\rho(\xi) = \frac{1}{Z} \; \exp\{ - \bt H(\xi) \} \; ,
\ee
with the partition function
\be
\label{18}
Z =  {\rm Tr} \int \exp\{ - \bt H(\xi) \} \; \cD\xi \;   ,
\ee
and with the grand Hamiltonian
\be
\label{19}
H(\xi) = \bigoplus_f H_f(\xi_f) \; , \qquad
H_f(\xi_f) = \hat H_f(\xi_f) - \sum_i \mu_i \hat C_{fi}(\xi_f) \; .
\ee

In the second quantization representation, taking into account the identity
$$
\int_{V_f} d\br = \int_V \xi_f(\br) \; d\br \; ,
$$
the partial Hamiltonians are
$$
\hat H_f(\xi_f) = \int \xi_f(\br) \; \psi_f^\dgr(\br) \;
\left( - \; \frac{\nabla^2}{2m} \right ) \; \psi_f(\br) \; d\br +
$$
\be
\label{20}
 + \frac{1}{2} 
\int \xi_f(\br) \; \xi_f(\br') \; \psi_f^\dgr(\br) \; \psi_f^\dgr(\br') \; 
\Phi(\br-\br')\; \psi_f(\br') \; \psi_f(\br) \; d\br d\br' \; .
\ee
Here and in what follows, where the spatial volume of integration is not specified, 
it is assumed to be over the whole system volume $V$. Similarly, the operators of 
observables have the form
$$
\hat A(\xi) = \bigoplus_f \hat A_f(\xi_f) \; ,
$$
\be
\label{21}
\hat A_f(\xi_f) = \sum_m \int \xi_f(\br_1) \xi_f(\br_2) \ldots \xi_f(\br_m)\;
A_f(\br_1,\br_2,\ldots,\br_m) \; d\br_1 d\br_2 \ldots d\br_m \; .
\ee
The observable quantities are given by the averages
\be
\label{22}
 \lgl \; \hat A \; \rgl = 
{\rm Tr} \int \hat\rho(\xi) \;\hat A(\xi) \; \cD\xi \;  .
\ee
The grand thermodynamic potential is
\be
\label{23}
\Om = - T \ln Z = - 
T \ln {\rm Tr} \int \exp\{-\bt H(\xi) \} \; \cD\xi \;  .
\ee

In order to realize practical calculations, it is necessary to explicitly 
define the procedure of averaging over phase configurations. To this end, let 
us introduce the variable
\be
\label{24}
 x_f \equiv \frac{1}{V} \int \xi_f(\br) \; d\br  
\ee
normalized as
\be
\label{25}
\sum_f x_f = 1 \;, \qquad 0 \leq x_f \leq 1 \;  .
\ee
Then the differential measure for the functional integration over manifold indicator 
functions is defined as
\be
\label{26}
 \cD \xi = \lim_{ \{\nu_f\ra\infty\} } \dlt\left( \sum_f x_f -1 \right)
\prod_f \prod_{j=1}^{\nu_f} \frac{d\ba_{fj}}{V} \; d x_f \;  ,
\ee
where the limit (\ref{8}) is assumed, ${\bf a}_{fj} \in V$ and $x_f\in [0,1]$, in
agreement with normalization (\ref{25}).

It is convenient to define an effective renormalized grand Hamiltonian by the relation
\be
\label{27}
 \exp(-\bt \widetilde H ) = \int \exp\{ - \bt H(\xi) \} \; \cD\xi \;  .
\ee
The following steps are based on the theorem formulated below.

\vskip 2mm

{\bf Theorem}. Consider the grand thermodynamic potential (\ref{23}), with the grand 
Hamiltonian defined in Eqs. (\ref{19}) and (\ref{20}). Accomplishing the averaging 
over configurations, implying the functional integration over the manifold indicator 
functions with the differential measure (\ref{26}), yields the grand thermodynamic 
potential 
\be
\label{28}
\Om = - T \ln {\rm Tr} \exp( - \bt \widetilde H ) \;  ,
\ee
with the renormalized grand Hamiltonian
$$
\widetilde H = \bigoplus_f H_f(w_f) \; , \qquad  
H_f(w_f) = \hat H_f(w_f) - \sum_i \mu_i \hat C_{fi}(w_f) \; ,
$$
$$
\hat H_f(w_f) = 
w_f \int \psi_f^\dgr(\br) \; \left( - \; 
\frac{\nabla^2}{2m}\right) \; \psi_f(\br) \; d\br +
$$
\be
\label{29}
 + \frac{1}{2} \; w_f^2 \int \psi_f^\dgr(\br) \; \psi_f^\dgr(\br') \;
\Phi(\br-\br') \; \psi_f(\br') \; \psi_f(\br) \; d\br d\br' \; ,
\ee
and where $w_f=V_f/V$ is defined as a minimizer of the grand thermodynamic potential
\be
\label{30}
\Om = {\rm abs} \; \min \sum_f \Om_f(w_f) \; , \qquad
\Om_f(w_f) = - T \ln {\rm Tr}_{\cH_f} \exp\{ - \bt H_f(w_f) \} \; ,
\ee
under the normalization condition
\be
\label{31}
 \sum_f w_f = 1 \; , \qquad 0 \leq w_f \leq 1 \;  .
\ee

The observable quantities (\ref{22}) become
$$
\lgl \;\hat A\; \rgl = {\rm Tr} \;\hat\rho\; \hat A = \sum_f 
\lgl \;\hat A_f \; \rgl \; , 
\qquad
\lgl \;\hat A_f \; \rgl = 
{\rm Tr}_{\cH_f} \;\hat\rho_f(w_f) \; \hat A_f(w_f) \; ,
$$
\be
\label{32}
\hat A_f(w_f) = \sum_m w_f^m 
\int A_f(\br_1,\br_2,\ldots,\br_m) \; d\br_1 d\br_2 \ldots d\br_m \;   ,
\ee
with the renormalized statistical operator
\be
\label{33}
\hat\rho = \frac{1}{Z} \; 
\exp( -\bt \widetilde H ) = \prod_f \hat\rho_f(w_f) \; ,
\qquad  
\hat\rho_f(w_f) = \frac{1}{Z_f} \; 
\exp\{ -\bt H_f(w_f) \} \; ,
\ee
in which
\be
\label{34}
 Z = {\rm Tr} \exp( -\bt \widetilde H ) = \prod_f Z_f \; ,
\qquad
Z_f = {\rm Tr}_{\cH_f} \exp\{-\bt  H_f(w_f) \} \;  .
\ee

\vskip 2mm

{\it Proof}. The proof of the theorem, with all mathematical details, has been 
thoroughly expounded in Refs. 
\cite{Yukalov_24,Yukalov_25,Yukalov_35,Yukalov_36,Yukalov_37,Yukalov_38}. The basic 
points of the proof are as follows. According to the definition for the function
of operators, the exponentials of Hamiltonians (\ref{20}) are expanded in powers
of the Hamiltonians, which leads to the functional polynomials in powers of the 
manifold indicator functions (\ref{2}). Each indicator function (\ref{2}) is written 
as the sum (\ref{7}) of the submanifold indicator functions that in $d$-dimensional 
space have the form
$$
\xi_{fj}(\br-\ba_{fj}) = \prod_{\al=1}^d   \xi_{fj}(r_\al-a_{fj}^\al)
$$ 
of the product of single-dimensional indicator functions
\begin{eqnarray}
\nonumber
\xi_{fj}(r_\al-a_{fj}^\al) = \left\{ \begin{array}{ll}
1 , ~ & ~ a_{fj}^\al - b_\al < r_\al < a_{fj}^\al + b_\al \\
0 , ~ & ~ r_\al < a_{fj}^\al - b_\al \; , ~ r_\al > a_{fj}^\al + b_\al 
\end{array} 
\right. \; ,
\end{eqnarray}
with $b_\alpha$ being the cell half-length in the direction $\alpha$. The 
single-dimensional indicator functions can be written in the Dirichlet 
representation as
$$
\xi_{fj}(r_\al-a_{fj}^\al) = \frac{1}{\pi} \int_{-\infty}^\infty
\frac{\sin(b_\al z) }{z} \; \exp\{ i z (r_\al - a_{fj}^\al) \} \; dz \; .
$$   
Then it is possible to directly integrate over the variables $a_{fj}^\al$, 
as is required in the definition of the differential measure (\ref{26}). 
After this integration, the resulting series is exponentiated leading to 
the form of the effective renormalized Hamiltonian (\ref{29}). 

\vskip 2mm

In this way, after the averaging over randomly distributed phase configurations, 
the problem reduces to the copies of the system corresponding to different phases, 
with renormalized Hamiltonians.

It is worth emphasizing that the coexisting phases are interdependent, since their 
effective Hamiltonians, resulting from the averaging over phase configurations, are 
renormalized by means of the phase probabilities $w_f$ satisfying conditions (\ref{31}). 
The phase probabilities are to be found from the minimization of thermodynamic 
potential.  

Different regions of the system are in chemical equilibrium with each other. This 
implies that the regions are correlated with each other through particle exchange. 
This correlation is taken into account by the standard for equilibrium statistical 
states equality of chemical potentials of different phases, that is the chemical 
potentials of the solid and liquid phases.

\section{Superfluid solid}

Let us now specify the problem by considering a solid with regions of disorder, 
such as dislocations or grain boundaries, in the cores of which there can exist 
disordered embryos of a liquid-like phase. The regions of disorder are randomly 
distributed over the sample. After averaging over phase configurations, as described 
in the previous section, we come to a renormalized Hamiltonian describing 
coexisting solid-like and liquid-like phases. 

A solid with superfluid properties is often named ``supersolid". However this term 
does not seem to be grammatically accurate. The standard meaning of the word ``super" 
accentuates the given property, but does not contradict it. For instance,
``superradiance" means superstrong radiance. ``Superconductivity" implies superstrong 
conductivity. ``Superfluidity" signifies superstrong fluidity. Therefore ``supersolidity",
according to the rules of grammatics, should characterize superrigid solidity. 
Contrary to this, vice versa, one talks not about a superrigid solid but about a solid 
with liquid superfluid properties. Hence this makes the term grammatically confusing.  
In addition, the term ``supersolid" has already been used for many years with respect 
to crystals in space dimensionality larger than three \cite{Oxford}.

\subsection{General relations}

The total number of particles $N$ in the volume $V$ is composed of two particle 
species forming a solid-like phase of $N_{sol}$ particles in a volume $V_{sol}$ 
and a liquid-like phase of $N_{liq}$ particles randomly distributed in a volume 
$V_{liq}$, such that
\be
\label{35}   
 N_{sol}+ N_{liq}= N \; , \qquad V_{sol} + V_{liq} = V \;  .
\ee
The corresponding geometric weights of the phases are
\be
\label{36}
 w_{sol} \equiv \frac{V_{sol}}{V} \; , \qquad
w_{liq} \equiv \frac{V_{liq}}{V} \;  .
\ee
It is also possible to introduce the particle fractions
\be
\label{37}
 n_{sol} \equiv \frac{N_{sol}}{N} \; , \qquad
n_{liq} \equiv \frac{N_{liq}}{N} \;  
\ee
and the phase densities
\be
\label{38}
\rho_{sol} \equiv \frac{N_{sol}}{V_{sol}} \; , \qquad
\rho_{liq} \equiv \frac{N_{liq}}{V_{liq}} \;   .
\ee
The overall average density is
\be
\label{39}
 \rho \equiv \frac{N}{V} = w_{sol} \rho_{sol} + 
w_{liq} \rho_{liq} \;  .
\ee

Usually, the density of a solid phase does not differ much from that of a 
liquid phase under the same conditions. In that case, the probabilities and 
fractions of a phase coincide,
\be
\label{40}
w_{sol} = n_{sol} \; , \qquad w_{liq} = n_{liq} \qquad
 ( \rho_{sol} = \rho_{liq} = \rho ) \; .
\ee

If the liquid-like phase allows for Bose-Einstein condensation, then among 
the particles of that phase there are $N_0$ Bose condensed particles and 
$N_1$ uncondensed particles,
\be
\label{41}
N_{liq} = N_0 +N_1 \;   .
\ee
Respectively, it is straightforward to define the related densities
\be
\label{42}
\rho_0 \equiv \frac{N_0}{V_{liq} } \; , \qquad 
\rho_1 \equiv \frac{N_1}{V_{liq} } \; ,
\ee
and fractions
\be
\label{43}
n_0  \equiv \frac{N_0}{N_{liq} } = \frac{\rho_0}{\rho_{liq} } \; ,
\qquad
n_1  \equiv \frac{N_1}{N_{liq} } = \frac{\rho_1}{\rho_{liq} } \; .
\ee
Similarly, one can define the fractions with respect to the total number of 
particles in the whole system,
\be
\label{44}
\overline n_0 \equiv \frac{N_0}{N} = n_{liq} n_0 \; , 
\qquad
\overline n_1 \equiv \frac{N_1}{N} = n_{liq} n_1 \;   .
\ee
The normalization conditions 
\be
\label{45}
\rho_0 + \rho_1 = \rho_{liq} \; , \qquad n_0 +n_1 = 1\; ,
\qquad
\overline n_0 + \overline n_1 = n_{liq}  
\ee
are valid.

The renormalized grand Hamiltonian is
\be
\label{46}
\widetilde H = H_{sol} \bigoplus H_{liq} \; ,
\ee
where $H_{sol}$ corresponds to a solid-state phase, while $H_{liq}$, to a 
liquid-like phase.

The general form of the Hamiltonian in the second quantization representation 
is actually the same for any system of particles with two-body interactions. 
As is well known, the system of particles with the same two-body interaction 
potential can form different thermodynamic phases. Mathematically, the difference 
arises through the features of the Fock spaces which the Hamiltonians are defined 
on. Each Fock space is formed by microstates possessing the properties typical 
of the considered phase, such as symmetry. The standard procedure of selecting 
the states with the required symmetry is realized by imposing constraints on the 
averages characterizing the related phases. Thus the solid phase is defined so 
that the average particle density be periodic over a crystalline lattice with a 
lattice vector $\ba$, 
$$
\lgl \; \psi_{sol}^\dgr(\br + \ba) \; \psi_{sol}(\br + \ba) \; \rgl =
\lgl \; \psi_{sol}^\dgr(\br) \; \psi_{sol}(\br) \; \rgl  \; ,
$$  
while the density of the liquid-like phase has to be uniform,
$$
 \lgl \; \psi_{liq}^\dgr(\br) \; \psi_{liq}(\br) \; \rgl =
\lgl \; \psi_{liq}^\dgr(0) \; \psi_{liq}(0) \; \rgl  \;  .
$$
If the liquid phase can display Bose-Einstein condensation, then its space of 
microstates needs to have broken global gauge symmetry, contrary to the solid 
state where the gauge symmetry is not broken, so that
$$
 \lgl \; \psi_{sol}(\br) \; \rgl = 0 \; , \qquad
 \lgl \; \psi_{liq}(\br) \; \rgl \neq 0 \;   .
$$
The averages complimented with the conditions selecting the required symmetry 
properties are often called {\it quasi-averages} \cite{Bogolubov_17,Bogolubov_18}.

\subsection{Solid-state phase}

The grand Hamiltonian of the solid-state phase is
\be
\label{47}
H_{sol} = \hat H_{sol} - \mu \hat N_{sol} \;   ,
\ee
with the energy Hamiltonian
$$
\hat H_{sol} = w_{sol} \int \psi_{sol}^\dgr(\br) \; \left( - \; 
\frac{\nabla^2}{2m} \right) \; \psi_{sol}(\br) \; d\br \; +
$$
\be
\label{48}
+ \;
\frac{1}{2} \; w_{sol}^2 \int \psi_{sol}^\dgr(\br) \; \psi_{sol}^\dgr(\br') \;
\Phi(\br-\br') \; \psi_{sol}(\br') \; \psi_{sol}(\br) \; d\br d\br'
\ee
and the number-of-particle operator
\be
\label{49}
 \hat N_{sol} = 
w_{sol} \int  \psi_{sol}^\dgr(\br) \; \psi_{sol}(\br) \; d\br \; .
\ee

There is the well known problem related to the fact that the interaction potential 
$\Phi({\bf r})$ can be nonintegrable and leading to divergences. It is also known 
that the way of avoiding this problem is to take account of short-range particle 
correlations, as a result of which, instead of the bare potential $\Phi({\bf r})$, 
there appears the correlated potential
\be
\label{50} 
 \widetilde\Phi(\br) = g(\br) \; \Phi(\br) \;  .   
\ee
The short-range correlation function $g({\bf r})$ smooths the interaction potential 
so that the correlated potential becomes integrable. The Hartree approximation with 
the correlated potential has been suggested by Kirkwood \cite{Kirkwood_39}. It has 
been proved \cite{Yukalov_40} that, starting from the Kirkwood approximation, it is 
possible to develop an iterative procedure containing in all orders only the 
correlated potential and having no divergences. 
   
In the present paper, our main aim is to find out whether superfluidity can happen 
in quantum crystals with regions of disorder. Since, most probably, this phenomenon 
happens at low temperatures, we consider the case of $T = 0$. Then free energy 
coincides with internal energy 
\be
\label{51}
 E_{sol} = \frac{1}{N} \; \lgl \; \hat H_{sol} \; \rgl \;  .
\ee
Quantum crystals are well described by the self-consistent harmonic approximation
\cite{Guyer_41,Yukalov_42,Yukalov_43,Yukalov_44}, which we use here. Following the 
standard approach, we expand the field operators in well localized Wannier functions
\cite{Marzari_46} and then expand the effective interactions in powers of deviations 
from the lattice sites up to second order. In the self-consistent harmonic 
approximation at zero temperature, we find the energy of the solid state 
\be
\label{52}
 E_{sol} = \frac{\rho_{sol}}{\rho} \; \left( \frac{1}{2}\; w_{sol}^2 u_0 +
\frac{9}{8} \; w_{sol}^{3/2} T_D \right) \qquad ( T = 0 ) \; ,
\ee
where 
\be
\label{53}
u_0 = \frac{\rho_{sol}}{\rho} \sum_j \widetilde\Phi(\ba_j) \qquad
(\ba_j \neq 0 )
\ee
is the potential well at a lattice site and 
\be
\label{54}
T_D = \left[ \; \frac{2\rho_{sol}}{3m\rho} \sum_j \; \sum_\al
\frac{\prt^2\widetilde\Phi(\ba_j)}{\prt a_j^\al \prt a_j^\al} \; \right]^{1/2}
\ee
is the Debye temperature.

At zero temperature, the lattice-site potential well $u_0$ is connected with the 
configurational potential energy per particle $U_{sol}$ of an ideal crystal through 
the relation
\be
\label{55}
 U_{sol} \equiv \frac{1}{2N} \sum_{i\neq j} 
\widetilde\Phi(\ba_i - \ba_j)  = \frac{1}{2}\; u_0 \; .
\ee

Note that expression (\ref{52}) for the energy of the solid state differs from the 
energy in the usual self-consistent harmonic approximation by the renormalization
due to the geometric probability of the solid state $w_{sol}$.

\subsection{Liquid-like phase}

If the liquid-like phase can exhibit Bose-Einstein condensation, then the global gauge 
symmetry must be broken. The symmetry breaking is realized by the Bogolubov shift 
\cite{Bogolubov_17,Bogolubov_18} of the field operator
\be
\label{56}
 \psi_{liq}(\br) = \eta(\br) + \psi_1(\br) \; ,
\ee
where the condensate function $\eta$ plays the role of an order parameter
\be
\label{57}
\eta(\br) \equiv \lgl \; \psi_{liq}(\br) \; \rgl \;   ,
\ee
while the second term describes a field operator of uncondensed particles, such that
\be
\label{58}
 \lgl \; \psi_1(\br) \; \rgl = 0 \; .
\ee
The latter condition conserves quantum numbers associated with the system particles, 
for instance momentum. 

In order to avoid double counting of degrees of freedom, the condensate function and 
the field operator of uncondensed particles are assumed to be orthogonal,
\be
\label{59}
\int \eta^*(\br) \; \psi_1(\br) \; d\br = 0 \; .
\ee
Then the number-of-particle operator of the liquid-like phase reads as the sum
\be
\label{60}
\hat N_{liq} = 
w_{liq} \int \psi_{liq}^\dgr(\br) \; \psi_{liq}(\br) \; d\br =
N_0 + \hat N_1 
\ee
of the number of condensed particles 
\be
\label{61}
N_0 = w_{liq} \int |\; \eta(\br) \; |^2 d\br = 
w_{liq} \int |\; \lgl \; \psi_{liq}(\br) \; \rgl \; |^2 \; d\br
\ee
and of the number-of-particle operator for uncondensed particles
\be
\label{62}
 \hat N_1 = w_{liq} \int \psi_1^\dgr(\br) \;\psi_1(\br) \; d\br \; .
\ee 
Thus the total number of particles in the liquid-like phase, $N_{liq}$ is the sum of 
the number $N_0$ of condensed particles and the number 
\be
\label{63}
N_1 = \lgl \; \hat N_1 \; \rgl 
\ee
of uncondensed particles.

The grand Hamiltonian of the liquid phase takes the form
\be
\label{64}
 H_{liq} =\hat H_{liq} - \mu_0 N_0 - \mu_1 \hat N_1 - \hat\Lbd \; ,
\ee
in which the first term is the energy Hamiltonian
$$
\hat H_{liq} = w_{liq} \int \psi_{liq}^\dgr(\br) \; 
\left( -\;\frac{\nabla^2}{2m} \right)\; \psi_{liq}(\br) \; d\br \; +
$$
\be
\label{65}
+ \; 
\frac{1}{2} \; w_{liq}^2 
\int \psi_{liq}^\dgr(\br) \; \psi_{liq}^\dgr(\br') \;
\Phi(\br-\br') \; \psi_{liq}(\br') \; \psi_{liq}(\br) \; d\br d\br'   
\ee
and the other terms guarantee the validity of the Lagrange constraints, the 
normalization conditions (\ref{61}) and (\ref{63}), and the last term
\be
\label{66}
 \hat\Lbd = \int [\;\lbd(\br) \; \psi_1^\dgr(\br)  +
\lbd^*(\br) \; \psi_1(\br) \; ] \; d\br
\ee
guarantees the quantum-number conservation condition (\ref{58}). The chemical 
potential of the liquid-like phase coincides with that of the solid-like phase 
and is equal to
\be
\label{67}
 \mu = \mu_0 n_0 + \mu_1 n_1 \; .
\ee
The so defined grand Hamiltonian allows to develop a self-consistent theory of 
Bose-condensed systems, where the spectrum of excitations is gapless and all 
conservation laws are sustained \cite{Yukalov_47,Yukalov_48,Yukalov_49,Yukalov_50}. 

Calculating the reduced energy at zero temperature
\be
\label{68}
E_{liq} = \frac{1}{N} \; \lgl \; \hat H_{liq} \;\rgl \; ,
\ee 
we introduce the notation  
\be 
\label{69}
\Phi_0 \equiv \int \widetilde \Phi(\br) \; d\br =  4\pi \; \frac{a_s}{m}
\ee
for the interaction strength. We take into account that the average density of the 
solid phase is close to that of the liquid-like phase, setting $\rho_{liq}=\rho$. 
The ratio of the characteristic potential energy $\rho \Phi_0$ to the characteristic 
kinetic energy 
\be
\label{70}
E_K \equiv \frac{\rho^{2/3}}{2m}
\ee
defines the gas parameter
\be
\label{71}
\gm \equiv \rho^{1/3} \; a_s= \frac{\rho\Phi_0}{8\pi E_k} \; .
\ee

Employing the self-consistent Hartree-Fock-Bogolubov approximation, as in Refs. 
\cite{Yukalov_48,Yukalov_49,Yukalov_50}, we find the fraction of uncondensed 
particles
\be
\label{72}
 n_1 = \frac{s^3}{3\pi^2}\; w_{liq}^{3/2}  
\ee
and the condensate fraction
\be
\label{73}
n_0 = 1 \; - \; \frac{s^3}{3\pi^2}\; w_{liq}^{3/2} \; ,
\ee
with the dimensionless sound velocity $s$ satisfying the equation
\be
\label{74}
 s^2 = 4\pi \gm (n_0 + \sgm) \;  ,
\ee
where 
\be
\label{75}
\sgm \equiv \frac{1}{\rho} \; \lgl \; \psi_1(\br) \; \psi_1(\br) \; \rgl
\ee
is the anomalous average. For the latter, employing dimensional regularization at 
small interactions and analytical continuation to arbitrary interactions 
\cite{Yukalov_51}, we obtain
\be
\label{76}
\sgm = \frac{8}{\sqrt{\pi}} \; (\gm w_{liq} )^{3/2} \;
\left[\; n_0 + 
\frac{8}{\sqrt{\pi}} \; (\gm w_{liq} )^{3/2} \; \sqrt{n_0} \; \right] \; .
\ee
At zero temperature, the superfluid fraction equals that of the liquid phase $n_{liq}$.

The energy of the liquid-like phase at zero temperature, in terms of the characteristic 
energy (\ref{70}), is
\be
\label{77}
\frac{E_{liq}}{E_K} = \frac{16s^5}{15\pi^2}\; w_{liq}^{7/2} +
4\pi \gm w_{liq}^2 \left( 1 + n_1^2 - 2 n_1 \sgm - \sgm^2 \right) \; .
\ee

Again we see that the energy (\ref{77}) of the liquid-like Bose-condensed phase, arising 
in the regions of disorder inside a solid, and the energy of the pure liquid phase with 
Bose-Einstein condensate, found in the self-consistent Hartree-Fock-Bogolubov approximation 
in Refs. \cite{Yukalov_48,Yukalov_49,Yukalov_50,Yukalov_51} differ by the renormalization 
caused by the geometric probability $w_{liq}$.

\subsection{Numerical analysis}

The dimensionless energy of the crystal with regions of disorder, normalized to $E_K$, 
is
\be
\label{78}
 E = \frac{E_{sol}+E_{liq}}{E_K} \; .
\ee
It is convenient to introduce the dimensionless quantities for the depth of the 
lattice-site potential well
\be
\label{79}
u \equiv -\;\frac{u_0}{E_K}
\ee
and for the Debye temperature
\be
\label{80}
 t_D \equiv \frac{T_D}{E_K} \;  .
\ee
For brevity, let us use the notation
\be
\label{81}
 w_{sol} \equiv w \; , \qquad w_{liq} = 1 - w \;  .
\ee

The energy of a crystal with regions of disorder (\ref{78}) reads as
\be
\label{82}
 E = \frac{9}{8} \; w^{3/2} \; t_D -\; \frac{1}{2} \; w^2 u +
\frac{16s^5}{15\pi^2}\; ( 1 - w)^{7/2} +
4\pi\gm ( 1 - w )^2 \left( 1 +n_1^2 - 2n_1 \sgm - \sgm^2 \right) \; .
\ee
As is evident, this expression is not just a linear combination of the energies for solid 
and liquid phases, but the renormalized energy of a solid with randomly distributed regions 
of disorder filled by a liquid-like phase.  

The solid-state probability $w$ is defined as the minimizer of energy (\ref{82}). 
The latter has also to be compared with the energy of the system in the pure crystalline 
state, when $w=1$,
\be
\label{83}
E_{sol}^{all} = \lim_{w\ra 1} E = \frac{9}{8} \; t_D - \;\frac{1}{2} \; u \; ,
\ee
and with the energy of all the system being in the pure superfluid phase, when $w=0$,
\be
\label{84}
E_{liq}^{all} = \lim_{w\ra 0} E = \frac{16s^5}{15\pi^2} + 
4\pi\gm \left( 1 +n_1^2 - 2n_1 \sgm - \sgm^2 \right) \;  .
\ee

Being mostly interested whether superfluidity could appear in hcp $^4$He, we keep 
in mind the parameters corresponding to this solid. The hcp $^4$He contains a single 
atom in a lattice site, with $12$ nearest neighbors. The interaction between atoms 
can be described by the Aziz potential \cite{Aziz_52} that is often used. Actually, 
all we need is the value of the Debye temperature $T_D$, the depth of the potential 
well $u_0$, and the effective interaction strength $\Phi_0$. The properties of solid 
hcp $^4$He are well known from both experiment and numerical modeling, since this 
solid has been thoroughly studied by many authors
\cite{Hodgdon_53,Ceperley_54,Diallo_55,Maris_56,Cazorla_57,Vitiello_58,Chan_59,Cazorla_60}.
The accepted Debye temperature for the pressure 25.3 bars at zero temperature is 
$T_D = 25$ K. At this pressure and zero temperature, the density of solid $^4$He along 
the melting line is $\rho_{sol}=0.0288$ \AA$^{-3}$. The density of liquid $^4$He, under 
these conditions, along the freezing line is $\rho_{liq} = 0.0262$ \AA$^{-3}$. The ratio 
$\rho_{sol}/\rho_{liq} = 1.1$ shows that the density between solid and liquid states 
is not much different, hence it is admissible to set $\rho_{sol}=\rho_{liq}$. For the 
characteristic energy $E_K$ we have $E_K=0.572$ K. Then the Debye temperature in units 
of $E_K$ is $t_D=T_D/E_K=43.7$. For the scattering length $a_s=2.203$ \AA, the gas
parameter $\gm=0.677$. The configurational potential energy (\ref{55}) is $U_{sol}=-37.3$ K. 
Therefore the potential well is $u_0=-62.6$ K and $u\equiv|u_0|/E_K=109$. 

We compare the energy $E$ of the solid with superfluid regions of disorder, with the 
energy of the pure solid state $E_{sol}^{all}$ and the energy of the pure superfluid 
state $E_{liq}^{all}$. The probability of the solid state $w$ is defined as the minimizer 
of the energy $E$. Fixing the Debye temperature $t_D=43.7$ and the interaction 
strength $\gm=0.677$, we consider all quantities as functions of the potential well 
$u$. This analysis gives us the answers to two questions: (i) Is there a range of 
parameters where a quantum crystal could become superfluid due to the presence of 
regions of disorder? and (ii) Can the hcp $^4$He be such a solid with superfluid 
properties?  
       
Figure 1 shows that the superfluid solid can exist only for $u<75.577$. When the 
parameter $u$ increases from small values to $u_0=75.577$, at this point there happens
a first-order quantum phase transition from a superfluid solid to the usual solid. The 
superfluid solid can exist as a metastable system between $u_0=75.577$ and $u=76.39$.  

Figure 2 presents the condensate fraction $n_0=N_0/N_{liq}$ normalized to the number 
of particles in the liquid state and the condensate fraction $\overline{n}_0=N_0/N$ 
normalized to the total number of particles in the sample. The relation between these 
fractions is $\overline{n}_0=(1-w)n_0$.

The behavior of the solid-state probability $w$ is illustrated in Fig. 3. At the 
point $u_0=75.577$, the probability jumps to one and then only the usual solid can 
exist, while no superfluidity can happen. 

In Fig. 4, the normal average $n_1$ and the anomalous average $\sigma$ are shown. 
The normal average represents the fraction $n_1=N_1/N_{liq}$ of uncondensed particles 
in the liquid phase. The modulus of the anomalous average $|\sigma|$ describes the 
fraction of pair-correlated particles in the liquid-like phase. As is seen, the 
anomalous average is always larger than the normal average, hence it cannot be 
neglected.      

These results demonstrate that, in general, there exist parameters, where a crystalline 
solid with regions of disorder, such as dislocations, can exhibit superfluid properties. 
However for the parameters of hcp $^4$He, this model does not show the existence of 
superfluidity. 

It is necessary to keep in mind that the considered model gives the upper boundary 
for the  possible condensate fractions $n_0 = N_0/N_{liq}$ and, respectively, for 
$\overline{n}_0 = N_0/N$, with the sole constraint $\overline{n}_0 < n_0$. This is 
because, for simplicity, in the derivation of the model in the second section, the 
admissible fraction of the co-existing liquid-like phase $n_{liq}$ was assumed to 
be allowed for taking the values in the interval $[0,1]$. This, however, is a too 
wide range of allowed variation of $n_{liq}=N_{liq}/N$, since $n_{liq}$, in reality, 
is limited by the total possible fraction of particles in the regions of disorder. 
If, for concreteness, we consider dislocations representing the regions of disorder 
in solid hcp $^4$He, then we have to take into account the following limitation 
\cite{Souris_14}. The density of dislocations in hcp $^4$He is $10^4-10^6$ cm$^{-2}$, 
dislocation core radius is of order $10^{-7}$ cm, and dislocation spacing is 
$10^{-2}$ cm. Then the fraction of particles inside dislocations, with respect to the 
total number of particles, is of order $10^{-10}-10^{-8}$. Taking this into account, 
if the upper boundary of the condensate fraction in a solid is $\overline{n}_0=N_0/N$, 
then the realistic possible fraction of condensed particles in that solid is not more
than $\overline{n}_0  10^{-8}$. Respectively, if the upper boundary for a fraction is 
zero, the actual fraction is for sure zero.

\section{Conclusion}

We have developed a statistical model of a quantum crystal that can house regions 
of disorder exhibiting liquid-like properties. Examples of such regions of disorder
are dislocation networks and grain boundaries. The regions of disorder are randomly
distributed inside the sample, which requires to accomplish averaging over phase
configurations. As a result of the averaging, we derive a renormalized Hamiltonian
describing a crystal with regions of disorder. In the case of Bose particles, in the
liquid-like regions there can arise Bose-Einstein condensate, hence there can appear
superfluidity. The model allows for the direct evaluation of the probability and 
fraction of possible Bose condensate, which, of course, depends on the system 
parameters. For the parameters, characterizing solid hcp helium, the model does not
predict the existence of Bose condensate, although, in general, there is a range of 
parameters, where Bose-condensation and superfluidity inside the regions of disorder
in quantum crystals could arise.         

Generally, the theory is developed for any temperature. At finite temperatures, we 
need to consider the thermodynamic potential (30) that has the form
$\Omega = \Omega_{sol} + \Omega_{liq}$, where $\Omega_f$ are expressed through the 
renormalized Hamiltonians $H_f(w_f)$ defined in Eq. (29). The thermodynamic potential
$\Omega_{sol}$ can be easily calculated in the self-consistent harmonic 
approximation \cite{Guyer_41,Yukalov_42,Yukalov_43,Yukalov_44} and the thermodynamic
potential $\Omega_{liq}$ can be found in the self-consistent Hartree-Fock-Bogolubov 
approximation \cite{Yukalov_47,Yukalov_48,Yukalov_49,Yukalov_50,Yukalov_51}. 
 
In the main part of the paper, we concentrate on zero temperature because, first of all,
it is exactly at zero temperature where superfluidity, if any, can arise most probably 
and because presenting the general cumbersome formulas would essentially enlarge the 
length of the article surpassing the reasonable for a letter limit.

Since the main points of the theory are general, this approach can be applied to 
systems with different interaction potentials, for instance with dipolar forces. 
Different interaction potentials will lead to different values of the Debye temperature 
$T_D$, the depth of the potential well $u_0$, and the effective interaction strength 
$\Phi_0$. Respectively, depending on the system characteristics, superfluidity in the 
regions of disorder will either exist or not.

\vskip 3mm

{\bf CRediT authorship contribution statement}

\vskip 2mm

{\bf V.I. Yukalov}: Formal analysis, Methodology, Investigation, Writing – original draft, 
Visualization, Validation. {\bf E.P. Yukalova}: Formal analysis, Investigation, Writing – 
original draft, Visualization, Validation, Numerical calculations.

\vskip 3mm

{\bf Declaration of competing interest}

\vskip 2mm

The authors declare that they have no known competing financial interests or personal 
relationships that could have appeared to influence the work reported in this paper.

\vskip 3mm

{\bf Acknowledgements}

\vskip 2mm

This research did not receive any specific grant from funding agencies in the public, 
commercial, or not-for-profit sectors.

\newpage

\begin{center}
{\Large{\bf Figure Captions}}
\end{center}

\vskip 2cm
{\bf Figure 1}. The energy $E$ of a superfluid solid, as compared with the energy 
$E_{sol}^{all}$ of a pure crystalline state and the energy $E_{liq}^{all}$ of 
the pure liquid-like phase, as functions of the potential well depth at a 
lattice site $u$. The right-hand-side figure (b) shows in a larger scale the
region of intersection of $E$ and $E_{sol}^{all}$. 

\vskip 1cm
{\bf Figure 2}. The condensate fraction $n_0 = N_0/N_{liq}$ with respect to the number 
of particles in the liquid-like phase and the condensate fraction $\overline{n}_0=N_0/N$
with respect to the total number of particles in the system, as functions of $u$.

\vskip 1cm
{\bf Figure 3}. The probability of the solid state $w$ as a function of $u$. For $u > u_0$,
only the pure solid state can exist.

\vskip 1cm
{\bf Figure 4}. The dimensionless anomalous average $\sigma$ and the fraction of uncondensed 
particles in the liquid-like phase $n_1$, as functions of $u$.

\newpage

\begin{figure}[ht]
\centerline{
\hbox{ \includegraphics[width=8cm]{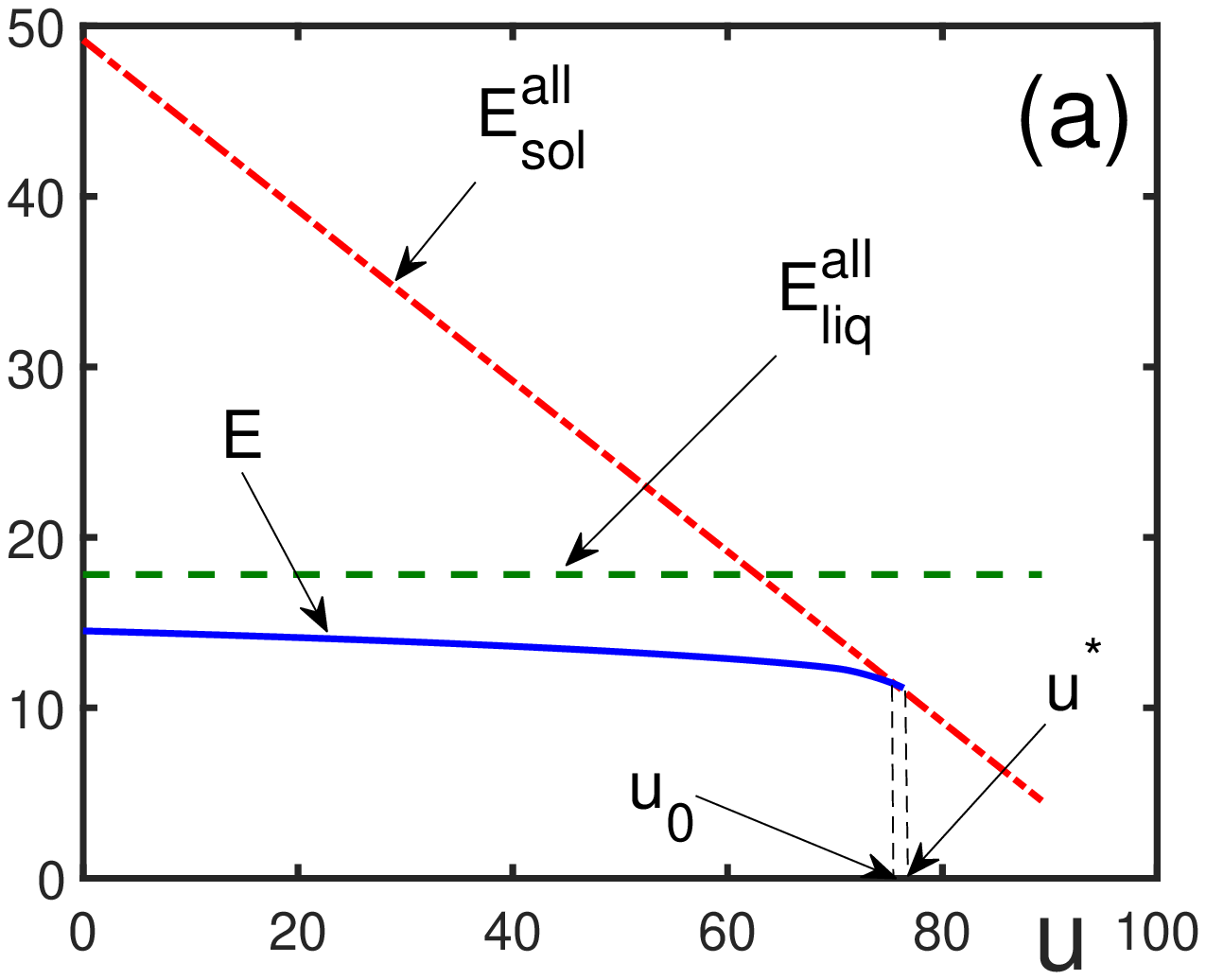} \hspace{0.5cm}
\includegraphics[width=8cm]{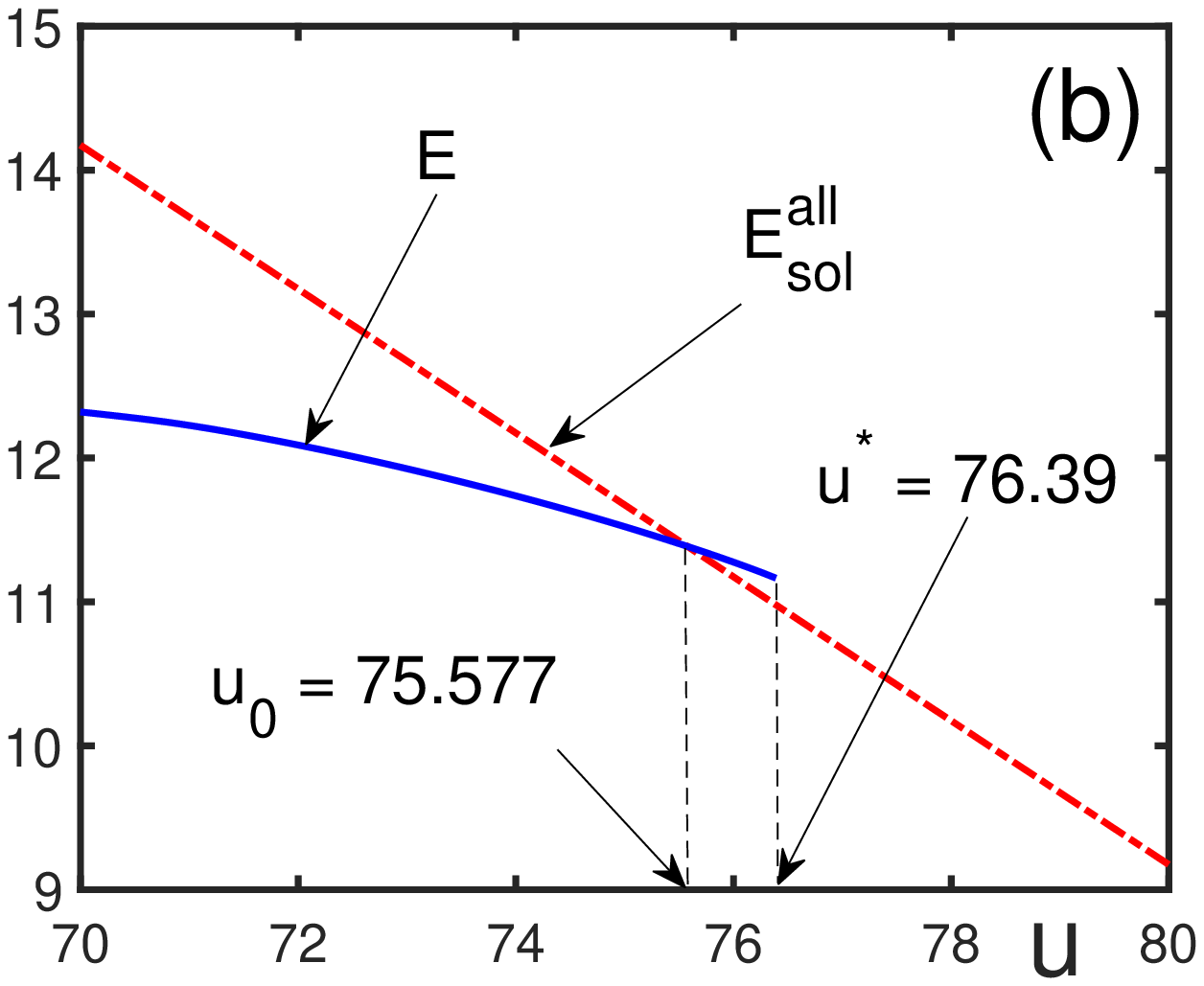}  } }
\caption*{\small {\bf Figure 1}.
The energy $E$ of a superfluid solid, as compared with the energy 
$E_{sol}^{all}$ of a pure crystalline state and the energy $E_{liq}^{all}$ of 
the pure liquid-like phase, as functions of the potential well depth at a 
lattice site $u$. The right-hand-side figure (b) shows in a larger scale the
region of intersection of $E$ and $E_{sol}^{all}$.
}
\end{figure}

\vskip 3cm
\begin{figure}[ht]
\centerline{
\includegraphics[width=8cm]{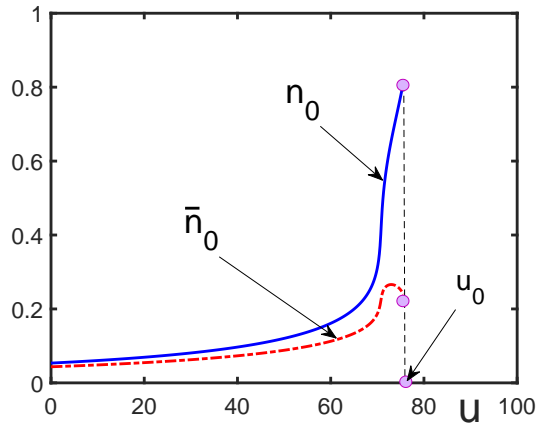}  }
\caption*{\small {\bf Figure 2}.
The condensate fraction $n_0 = N_0/N_{liq}$ with respect to the number 
of particles in the liquid-like phase and the condensate fraction $\overline{n}_0=N_0/N$
with respect to the total number of particles in the system, as functions of $u$.
}
\end{figure}

\vskip 3cm
\begin{figure}[ht]
\centerline{
\includegraphics[width=8cm]{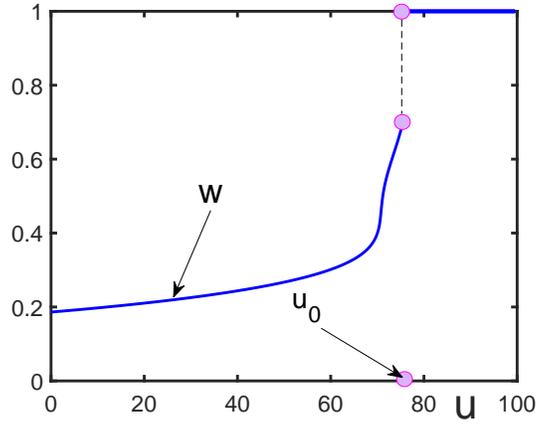}  }
\caption*{\small {\bf Figure 3}.
The probability of the solid state $w$ as a function of $u$. For $u > u_0$,
only the pure solid state can exist.
}
\end{figure}

\begin{figure}[ht]
\centerline{
\includegraphics[width=8cm]{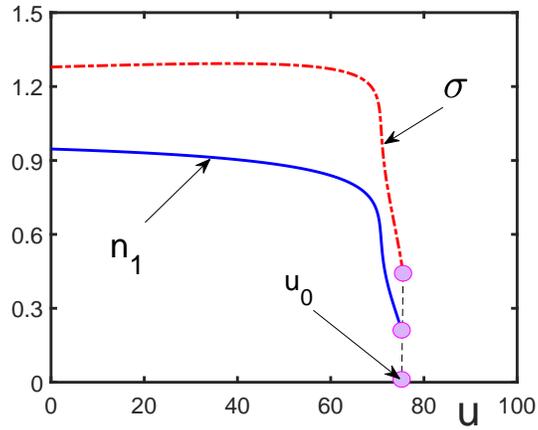} }
\caption*{\small {\bf Figure 4}.
The dimensionless anomalous average $\sigma$ and the fraction of uncondensed 
particles in the liquid-like phase $n_1$, as functions of $u$. 
}
\end{figure}

\end{document}